\documentclass[aps,prb,twocolumn,showpacs,floatfix]{revtex4}
\usepackage{graphics}
\usepackage{epsfig}
\usepackage{amsfonts}
\usepackage{amsmath}
\usepackage{amssymb}
\usepackage{pdfsync}
\usepackage[latin1]{inputenc}
\usepackage[thinspace,amssymb]{SIunits}
\begin{document}
\preprint{draft 1}
\title[ESR of Yb(Rh$_{1-x}$Co$_x$)$_2$Si$_2$]{Electron Spin Resonance of the Yb~$4f$ moment in Yb(Rh$_{1-x}$Co$_x$)$_2$Si$_2$}
\author{T. Gruner}
\author{J. Sichelschmidt}
\email{Sichelschmidt@cpfs.mpg.de}
\author{C. Klingner}
\author{C. Krellner}
\author{C. Geibel}
\author{F. Steglich}
\address{Max Planck Institute for Chemical Physics of Solids, D-01187 Dresden, Germany}
\email{Sichelschmidt@cpfs.mpg.de}

\date{\today}

\begin{abstract}
The evolution of spin dynamics from the quantum critical system YbRh$_{2}$Si$_{2}$ to the stable trivalent Yb system YbCo$_{2}$Si$_{2}$ was investigated by Electron Spin Resonance (ESR) spectroscopy. While the Kondo temperature changes by one order of magnitude, all compositions of the single crystalline series Yb(Rh$_{1-x}$Co$_x$)$_2$Si$_2$ show well defined ESR spectra with a clear Yb$^{3+}$ character for temperatures below $\approx 20$\,K.
With increasing Co-content the ESR $g$-factor along the $c$-direction strongly increases indicating a continuous change of the ground state wave function and, thus, a continuous change of the crystal electric field. The linewidth presents a complex dependence on the Co-content and is discussed in terms of the Co-doping dependence of the Kondo interaction, the magnetic anisotropy and the influence of ferromagnetic correlations between the 4$f$ states. The results provide evidence that, for low Co-doping, the Kondo interaction allows narrow ESR spectra despite the presence of a large magnetic anisotropy, whereas at high Co-concentrations, the linewidth is controlled by ferromagnetic correlations. A pronounced broadening due to critical correlations at low temperatures is only observed at the highest Co-content. This might be related to the presence of incommensurate magnetic fluctuations.  [published in Phys. Rev. B {\bf 85}, 035119 (2012)] 
\end{abstract}

\maketitle
\section{Introduction}
The heavy fermion metal YbRh$_{2}$Si$_{2}$ shows thermodynamic and transport properties which, at low temperatures, are determined by the hybridization of 4$f$ states and conduction electrons resulting in a Kondo interaction with a single-ion Kondo temperature $T_{\rm K}\simeq25$~K.\cite{gegenwart05a,kohler08a} Besides its quite remarkable magnetic behaviors, like weak antiferromagnetic (AFM) order below $T_{\rm N}=72$~mK, a field induced AFM quantum critical point at $B_{c}=60$~mT, and a highly enhanced Sommerfeld-Wilson ratio,\cite{gegenwart05a} the existence of a well-defined Electron Spin Resonance (ESR) signal at temperatures \textit{below} $T_{\rm K}$ with pronounced Yb$^{3+}$ Kondo-ion character triggered new theoretical studies of the dynamical properties of local and itinerant magnetism in Kondo lattice systems.\cite{schlottmann09a,wolfle09a,kochelaev09a,zvyagin09b} In these theories the origin of the resonance was intimately connected to the Kondo interaction and strong ferromagnetic correlations. The relevance of both effects were experimentally verified by ESR results on Lu-diluted YbRh$_{2}$Si$_{2}$,\cite{duque09a} ESR experiments under pressure and upon Co doping ($x\le0.18$),\cite{sichelschmidt10b} and the ESR of the ferromagnetic Kondo lattice CeRuPO.\cite{krellner08a,forster10a}\\
Recently the compound series Yb(Rh$_{1-x}$Co$_{x}$)$_{2}$Si$_{2}$ was investigated for the whole concentration range $0\le x\le1$. Whereas the crystal structure remains tetragonal (space group I4/mmm) the magnetic, thermal, and transport properties show clear variations with $x$:\cite{klingner11a,mufti10a,pedrero10a} it was shown that the Kondo interaction can largely be suppressed by substituting Rh with the smaller isoelectronic Co which, at the same time, enhances $T_{\rm N}$ and the size of the ordered moment, thus stabilizing the AFM order by suppressing the Kondo screening.\cite{klingner11a} For YbCo$_{2}$Si$_{2}$ with $T_{\rm N}=1.7$~K a stable trivalent Yb state with well defined crystal electric field (CEF) levels was evidenced by susceptibility \cite{kolenda89a}, M\"o\ss bauer results \cite{hodges87a}, inelastic neutron-scattering experiments \cite{goremychkin00a}, specific heat measurements \cite{klingner11b} as well as by photoemission spectroscopy \cite{klingner11a}. The Co-ion itself remains non-magnetic because of a strong Co-Si hybridization.\cite{kolenda89a} In the magnetically ordered state of YbCo$_{2}$Si$_{2}$ a complex magnetic phase diagram with significant basal-plane anisotropy is evidenced by the magnetoresistance \cite{mufti10a} and magnetization data \cite{pedrero11a}. Powder neutron diffraction revealed a complex magnetic structure with two phases with different propagation vectors.\cite{kaneko10a} A first-order phase transition from an incommensurable arrangement of moments below $T_{\rm N}$ into a commensurable one below $T_{\rm L}=0.9\,{\rm K}<T_{\rm N}$ was identified.\cite{pedrero10a} 
For YbRh$_{2}$Si$_{2}$ the stronger 4$f$-conduction electron hybridization leads to a small deviation of the valency from the trivalent state. Photoemission spectroscopy results suggested a mean valency of $\sim 2.9$.\cite{danzenbacher07a}
Yet, the magnetic structure of YbRh$_{2}$Si$_{2}$ is not known due to its very small ordered moment. Some hints could be obtained from the knowledge of the magnetic structure in YbCo$_{2}$Si$_{2}$ since within an isoelectronic series of compounds magnetic structures are often closely related.\\
In this paper we first present the ESR results of YbCo$_{2}$Si$_{2}$ and analyze the data in the context of further experimental results. We continue with the series Yb(Rh$_{1-x}$Co$_{x}$)$_{2}$Si$_{2}$ where we found a sensitive dependence of the ESR parameters on the Co -content. 
\section{Experimental Details}
\label{sec2}
For our ESR measurements we used clean In-flux grown single crystals of Yb(Rh$_{1-x}$Co$_{x}$)$_{2}$Si$_{2}$ which all had their $\boldsymbol{c}$-axis perpendicular to the main surface of the platelet-shaped crystals. Thermal and magnetic properties as well as electrical resistivity have been thoroughly investigated and were described elsewhere.\cite{klingner11a}\\ 
ESR detects the absorbed power $P$ of a magnetic microwave field $\boldsymbol{b_{\rm mw}}$ as a function of a transverse external static magnetic field $\boldsymbol{B}$. To improve the signal-to-noise ratio, a lock-in technique is used by modulating the static field, which yields the derivative of the resonance signal $dP/dB$. The ESR measurements were performed with a standard continuous wave spectrometer at X-band frequencies ($\nu \approx$ \unit{9.4}{\giga \hertz}) by using a cylindrical resonator in TE$_{012}$ mode or, for temperatures down to 1.5~K, a split-ring resonator. The temperature was varied between \unit{1.5}{\kelvin} $\leq T \leq$ \unit{16}{\kelvin} using $^4$He-flow-type cryostats. 
The ESR measurements at Q-band frequencies (34~GHz) turned out to provide much less definite spectra. Their linewidths and $g$-factors are found to be comparable to the X-band results regarding temperature dependencies and absolute values.\\
All recorded ESR spectra were analyzed by fitting them with a metallic Lorentzian line shape.\cite{wykhoff07b} From these fits the following ESR parameters were extracted: linewidth $\Delta B$ (HWHM), the ESR $g$-factor (as given by the resonance field $B_{res}$ in the resonance condition $h\nu = g \mu_B \cdot B_{res}$), the asymmetry parameter $\alpha$ (describing the dispersive contribution in metallic samples), and the intensity (spin susceptibility, given by the area $\propto amp\cdot\Delta B^{2}\cdot\sqrt{\alpha^{2}+1}$ under the ESR absorption).
%
%
\section{Experimental Results and Discussion}
\label{sec3}
The ESR resonance field displays for the whole series Yb(Rh$_{1-x}$Co$_x$)$_2$Si$_2$ ($x=0...1$) a pronounced anisotropy upon rotating the crystalline $c$-axis from parallel ($\Theta=0^\circ$) towards perpendicular ($\Theta=90^\circ$) to the magnetic field. This confirms the local character of the signal and verifies a typical resonance feature of Yb-ions which are located on lattice sites with tetragonal point symmetry.\cite{kutuzov08a,kutuzov11a} 
\subsection{YbCo$_{2}$Si$_{2}$}
\label{sec3A}
\begin{figure}[h]
 \centering
 \includegraphics[width=0.8\columnwidth]{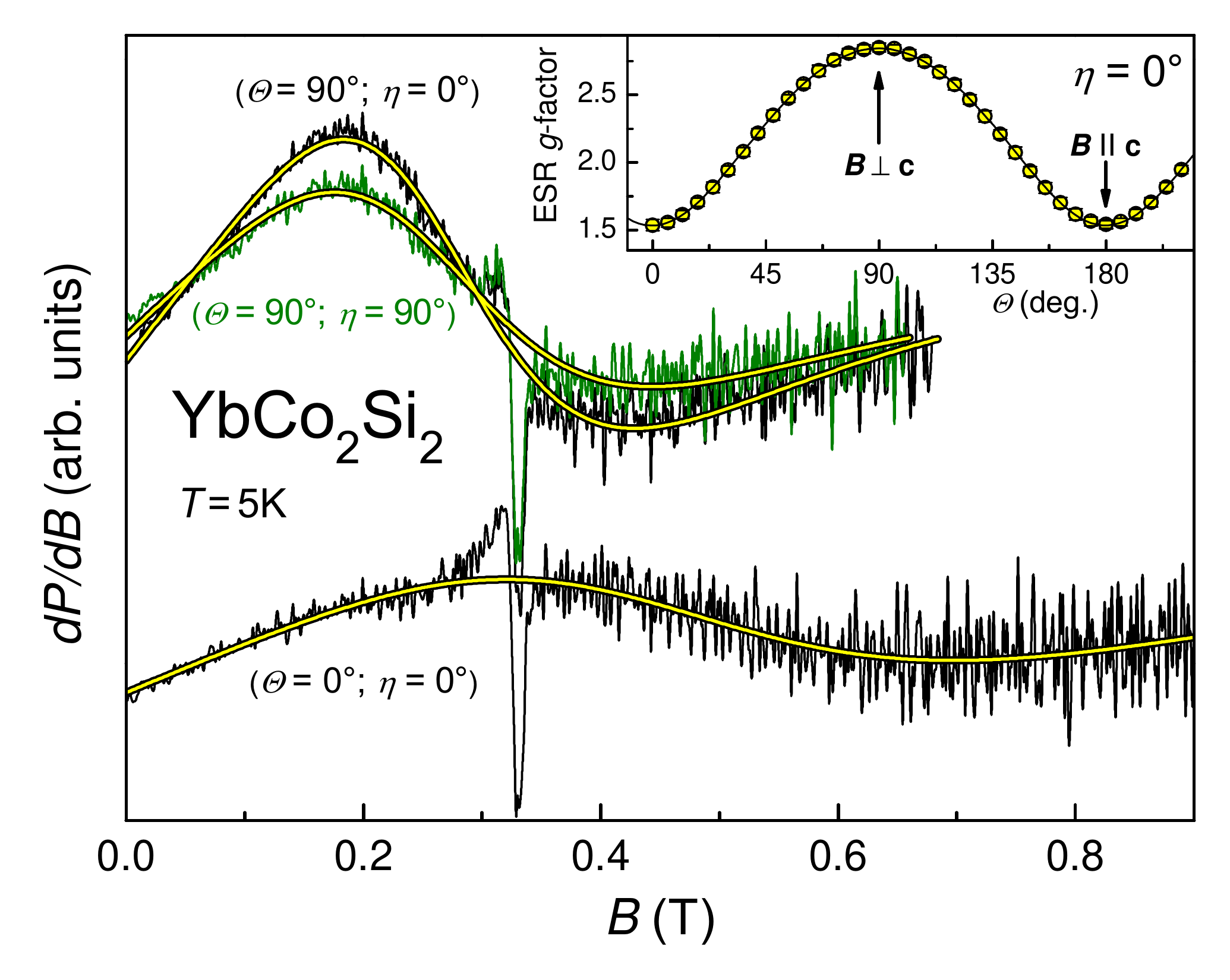}
  \caption{(color online) ESR spectra of YbCo$_{2}$Si$_{2}$ at $T=5$~K for the field orientation $\boldsymbol{B}\perp c$-axis ($\Theta=90^\circ$ and $\eta$-orientation of the basal plane to the microwave magnetic field) and $\boldsymbol{B}\,\|\, c$-axis ($\Theta=0^\circ$). Solid lines indicate Lorentzian lines ($\Theta=90^\circ$: same resonance field and linewidth for $\eta=0^\circ,90^\circ$). Inset shows the $\Theta$-dependence of the resonance-field-calculated $g$-factor with uniaxial behavior indicated by the solid line. Peak at 0.325~T arises from background signals of the resonator.} 
 \label{fig1}
\end{figure}

Fig.\ref{fig1} shows typical ESR spectra of YbCo$_{2}$Si$_{2}$ for selected directions of the static magnetic field $\boldsymbol{B}$ (angle $\Theta$) and the microwave magnetic field $\boldsymbol{b_{\rm mw}}$ (angle $\eta$). All spectra could be well described by a metallic Lorentzian lineshape (solid line). Within experimental accuracy linewidth and resonance field remain unchanged if the $c$-axis is rotated from $c \perp \boldsymbol{b_{\rm mw}}\; (\eta=0^\circ)$ to  $c\,\|\, \boldsymbol{b_{\rm mw}}\; (\eta=90^\circ)$. Interestingly, in that case, the ESR intensity only varies by a factor of 1.5. This is not expected from the considerable uniaxial anisotropy shown in the inserted graph for $c \perp \boldsymbol{b_{\rm mw}}$: With the observed values $g_{\bot}(\unit{5}{\kelvin})= 2.84\pm 0.03$ and $g_{\|}(\unit{5}{\kelvin})= 1.53\pm 0.02$ the intensity should vary by a factor of $(g_{\perp}/g_{\|})^{2}=3.4$. The origin of this discrepancy which is even more pronounced in YbIr$_{2}$Si$_{2}$ and YbRh$_{2}$Si$_{2}$ \cite{gruner10a} remains unclear.\\ 
%
\begin{figure}[h]
 \centering
 \includegraphics[width=0.8\columnwidth]{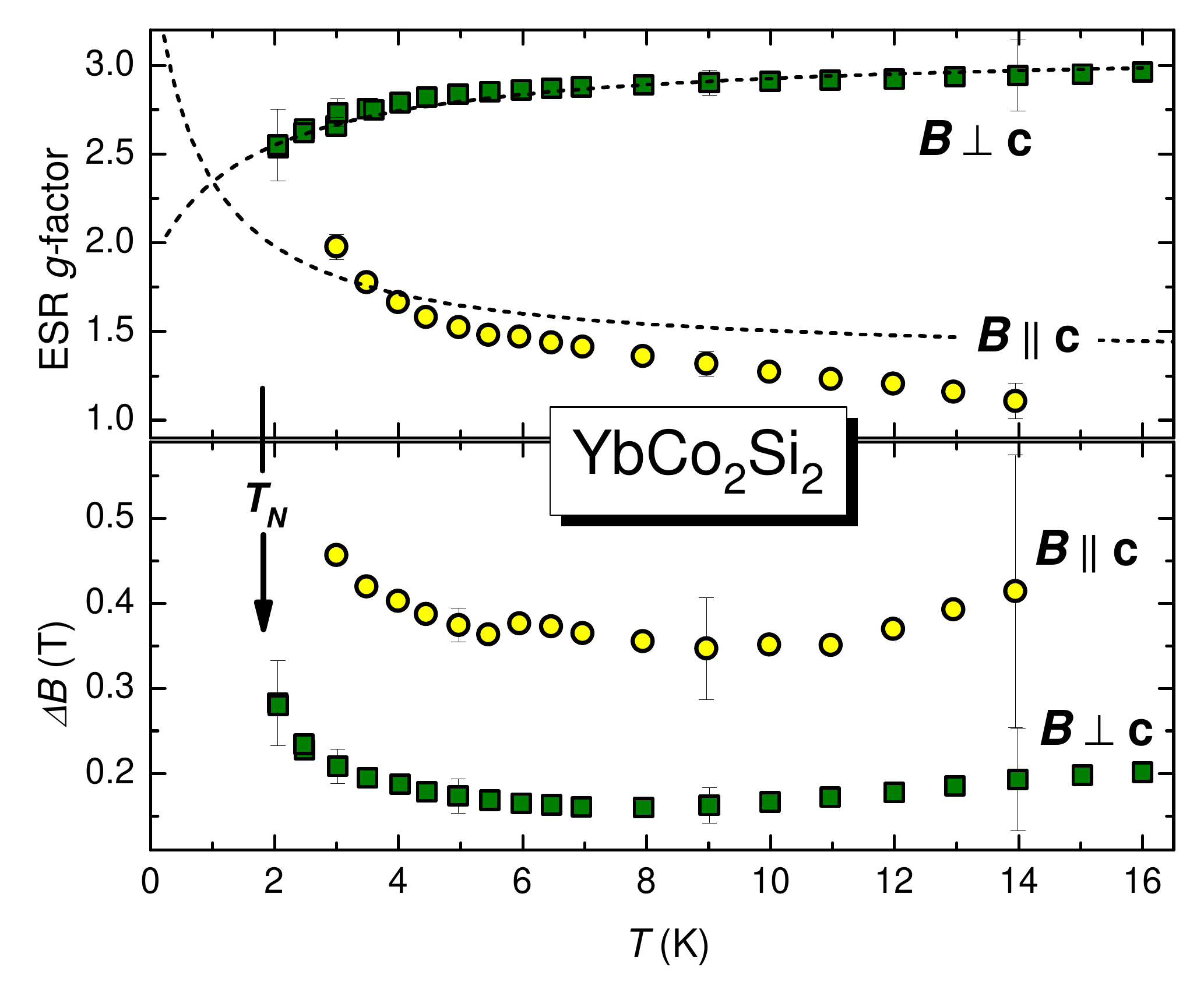}
  \caption{(color online) Temperature dependence of linewidth $\Delta B$ and ESR $g$-factor of YbCo$_{2}$Si$_{2}$ above the antiferromagnetic ordering temperature $T_{\rm N}$ for the field orientation $B\perp c$-axis and $B\,\|\, c$-axis. Dashed lines in $g(T)$ shows the mean field expectation as deduced from $\chi(T)$. Here, a larger crystal than the one of Fig. \ref{fig1} was used, hence yielding stronger lines and a higher data accuracy.} 
 \label{fig2}
\end{figure}
%
The observed $g$-values nicely agree with the values expected from the saturation magnetization \cite{pedrero11a}: $g_{\perp}^{m}=3.0$ ($\mu_{\perp}\simeq 1.5\mu_{\rm B}$) and $g_{\|}^{m}=1.2$ ($\mu_{\|}\simeq 0.6\mu_{\rm B}$). These $g$-factors can be well accounted for by a CEF model using a ground state doublet with a $\Gamma_{7}$ symmetry and a strong mixing of $|\pm3/2>$ and $|\mp5/2>$ wave functions.\cite{klingner11b} The deviations of the ESR $g$-values from $g^{m}$ are partly due to the pronounced temperature dependence of the $g$-values as shown in Fig.~\ref{fig2}. As a consequence of the anisotropic interactions between the Yb-ions the internal magnetic field leads to a temperature dependent shift of the resonance fields. This can reasonably be described by a molecular field model \cite{huber09a,huber76a}    
\begin{eqnarray}
\label{eq:gsenkT}
g_\bot(T) &=& g_{\bot}^0 \cdot \left(1-\frac{\theta^\| - \theta^\bot}{T - \theta^\bot}\right)^\frac{1}{2}\\
\label{eq:gparaT}
g_\|(T) &=& g_{\|}^0 \cdot \left(1+\frac{\theta^\| - \theta^\bot}{T - \theta^\|}\right)
\end{eqnarray}
as shown by the dashed lines in Fig.~\ref{fig2}. Here, the Weiss temperatures $\theta^{\bot,\|}$ are obtained by a Curie-Weiss [$\chi=C/(T-\theta)$] plot of the susceptibilities $\chi^{\perp,\|}$ below $T=6$~K measured at the resonance magnetic fields of 0.46~T and 0.35~T for $\chi^{\perp}$ and $\chi^{\|}$, respectively: $\theta^\bot = \unit{(-2.1 \pm 0.2)}{\kelvin}$ and $\theta^\| = \unit{(-0.8 \pm 0.2)}{\kelvin}$. Furthermore we used $g_{\bot}^0=g_{\perp}^{m}=2.96$ and $g_{\|}^0=g_{\|}^{m}=1.2$. Although the dashed lines in Fig.~\ref{fig2} do not agree perfectly with the data they, however, reflect the correct tendency. Namely, towards low temperatures the $g$-values tend to merge each other which is a consequence of $\theta^\| > \theta^\bot$. Note, in YbIr$_{2}$Si$_{2}$ $\theta^\| < \theta^\bot$ and thus we observed a mutual divergent behavior of $g_{\bot}(T)$ and $g_{\|}(T)$.\cite{gruner10a}\\ 
The temperature dependence of the ESR linewidth clearly differs at low temperatures from the linewidth behaviors of YbIr$_{2}$Si$_{2}$ and YbRh$_{2}$Si$_{2}$ where a continuous decrease or a saturation towards low temperatures was observed \cite{sichelschmidt07a,sichelschmidt10a}. As shown in Fig.~\ref{fig2} the linewidth of YbCo$_{2}$Si$_{2}$ starts to increase below $T\approx8$~K indicating the effect of critical spin fluctuations due to magnetic order below $T_{\rm N}\approx1.7$~K. Above $T\approx8$~K the linewidth continuously increases, similar to YbIr$_{2}$Si$_{2}$ and YbRh$_{2}$Si$_{2}$. For $T>16$~K the ESR signal is too weak and too broad to be detected. A comparison of X-band with Q-band data of YbCo$_{2}$Si$_{2}$ shows no change in linewidth and $g$-value regarding absolute value and temperature dependence.

\begin{figure}[t]
 \centering
 \includegraphics[width=0.7\columnwidth]{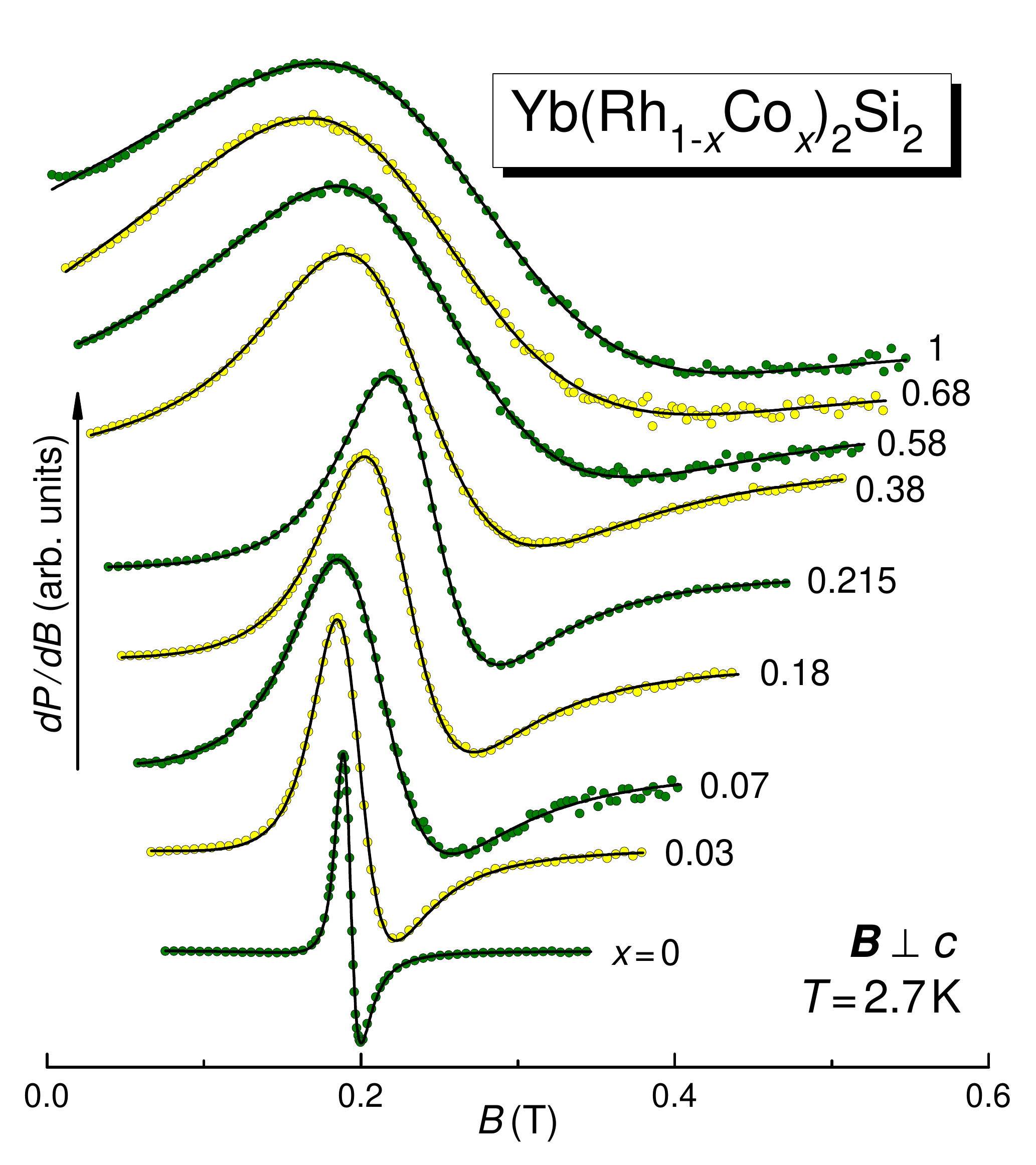}
  \caption{(color online) Evolution of the ESR spectra (symbols, background subtracted) by variation of the Co concentration for the field orientations $\boldsymbol{B}\perp c$ and $\boldsymbol{b_{\rm mw}} \perp c$. Solid lines depict metallic Lorentzian line shapes.} 
 \label{fig3}
\end{figure}

\subsection{Yb(Rh$_{1-x}$Co$_x$)$_2$Si$_2$}

Figure~\ref{fig3} presents the ESR signals of the series Yb(Rh$_{1-x}$Co$_x$)$_2$Si$_2$ at $T=2.7$~K and for the magnetic field $\boldsymbol{B}\perp c$-axis, i.e. the orientation at which the resonance field is the smallest. All depicted spectra show well defined metallic Lorentzian shapes. The ratio of dispersion to absorption varies between 0.64 and 1 without clear relation to the $x$ values which may be explained by experimental reasons like different resonator filling factors.\\ 
\begin{figure}[t]
 \centering
 \includegraphics[width=0.8\columnwidth]{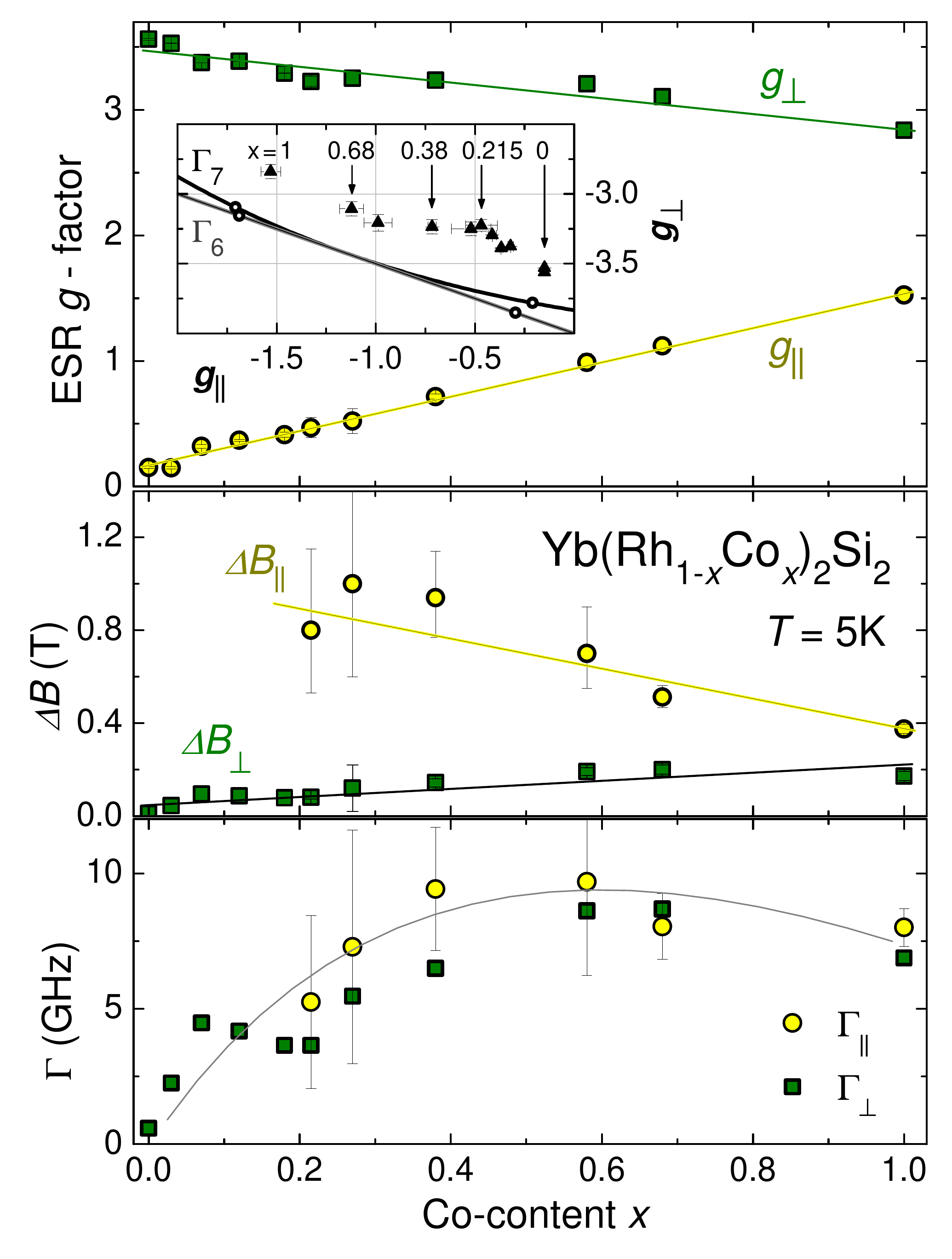}
  \caption{(color online) Co-concentration dependence of the ESR $g$-factor, linewidth $\Delta B$ and relaxation rate $\Gamma=\frac{\mu_{\rm B}}{h}g\Delta B$ for the field orientations $B\perp,\|\;c$-axis and for $T=5$~K. Solid lines are guides for the eyes. Insert: Evolution of $g$ factor anisotropy with Co-concentration in comparison with the expected $g$ factors (solid lines) which correspond to $\Gamma_{6}$ and $\Gamma_{7}$ groundstate symmetries for a Yb$^{3+}$ ion in a tetragonal crystal field.} 
 \label{fig4}
\end{figure}
%
Linewidth and resonance field show an overall continuous change with the Co-content $x$ as shown in Fig.~\ref{fig4} for $T=5$~K. The anisotropy of both quantities reduces with increasing $x$. Note, that $\Delta B_{\|}$ shows a narrowing with increasing Co-doping,
providing evidence that the relaxation due to disorder is not the dominant broadening mechanism. The relaxation rate which is determined by $\Gamma=\frac{\mu_{\rm B}}{h}g\Delta B$ shows no clear anisotropy but a remarkable increase with increasing $x$ up to $\approx0.5$ and a slight decreasing tendency with further increasing $x$ up to 1. Both $g_{\bot}$ and $g_{\|}$ evolve with $x$ with no pronounced anomalies, especially around $x=0.5$ where $T_{\rm N}$ shows a minimum.\cite{klingner11a}
The inset of Fig.~\ref{fig4} provides a comparison of the measured $g$-values at $T=5$~K (symbols) with those expected for the two possible Kramers doublets \cite{kutuzov08a,kutuzov11a} with irreducible representations $\Gamma_{6}$ or $\Gamma_{7}$. The experimental values are close to the theoretical values for either $\Gamma_{6}$ or $\Gamma_{7}$ doublets for all Co-contents. Thus, the ESR $g$-values can be well accounted for by a localized model. However, these results do not allow a clear distinction between $\Gamma_{6}$ and $\Gamma_{7}$.  
The latter was established for the ground state of YbRh$_{2}$Si$_{2}$ by angle-resolved photoemission investigations of the energy dispersion of the crystal-field-split 4$f$ states \cite{vyalikh10a} and by an analysis of the quadrupole moment \cite{geibel10a}. Furthermore, the analysis of all available data on YbCo$_{2}$Si$_{2}$ using a standard CEF model indicate a $\Gamma_{7}$ ground state for YbCo$_{2}$Si$_{2}$ too.\cite{klingner11b} All these results indicate a smooth and continuous evolution of the crystal electric field from YbRh$_{2}$Si$_{2}$ to YbCo$_{2}$Si$_{2}$ without change of the symmetry of the ground state doublet. 

\begin{figure}[t]
 \centering
 \includegraphics[width=1\columnwidth]{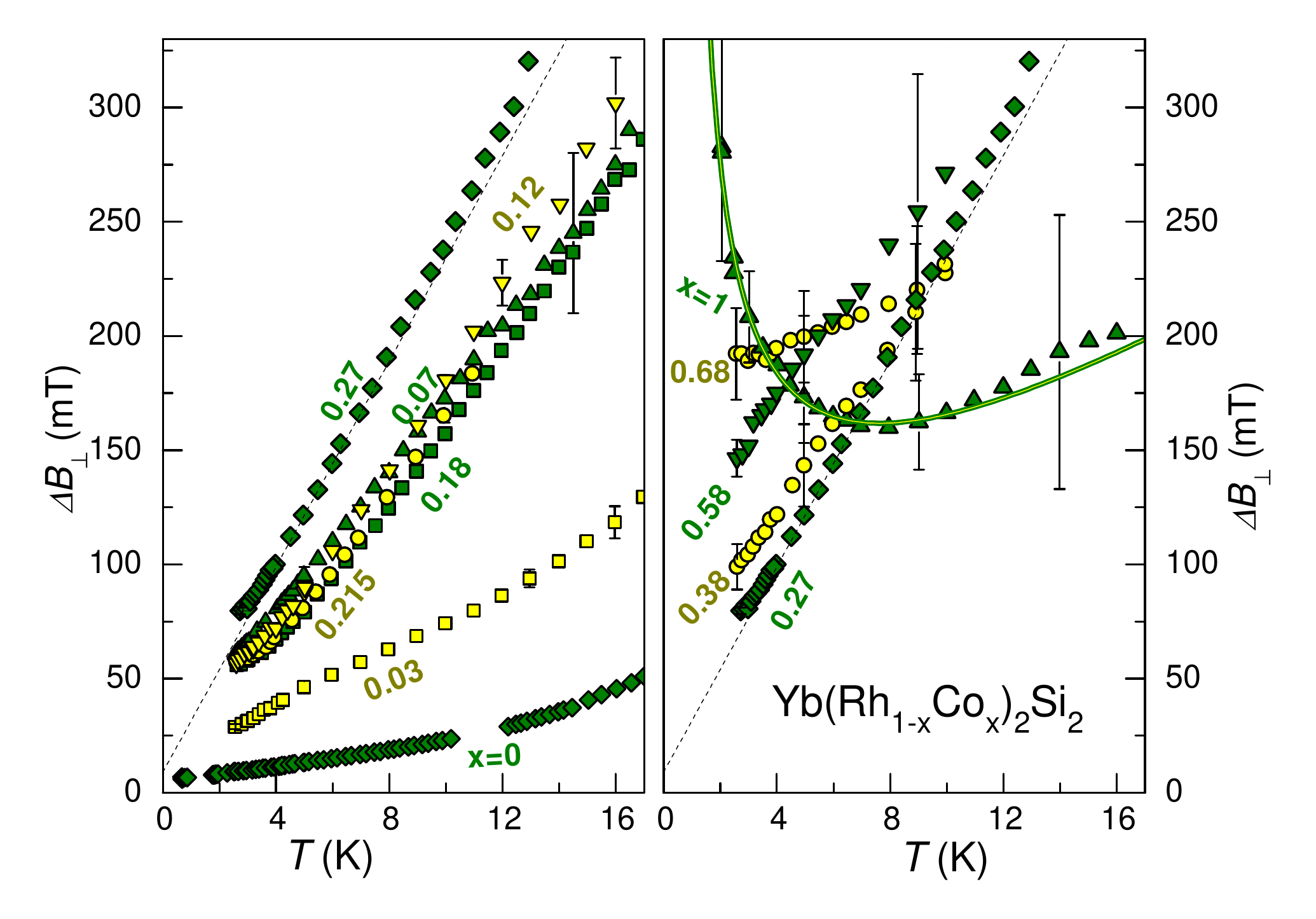}
  \caption{(color online) Temperature dependence of the linewidth $\Delta B$ for the field orientation $B\perp c$-axis. Dashed lines show for $x=0.27$ the maximum linear increase. The solid line describes the low temperature part of the $x=1$ data with a $1/\sqrt{T-T_{\rm L}}+\Delta B_{0}+b(T-T_{\rm L})$ behavior ($b=8$~mT/K), see main text Sec. \ref{disc}.} 
 \label{fig5}
\end{figure}

\begin{figure}[h]
 \centering
 \includegraphics[width=1\columnwidth]{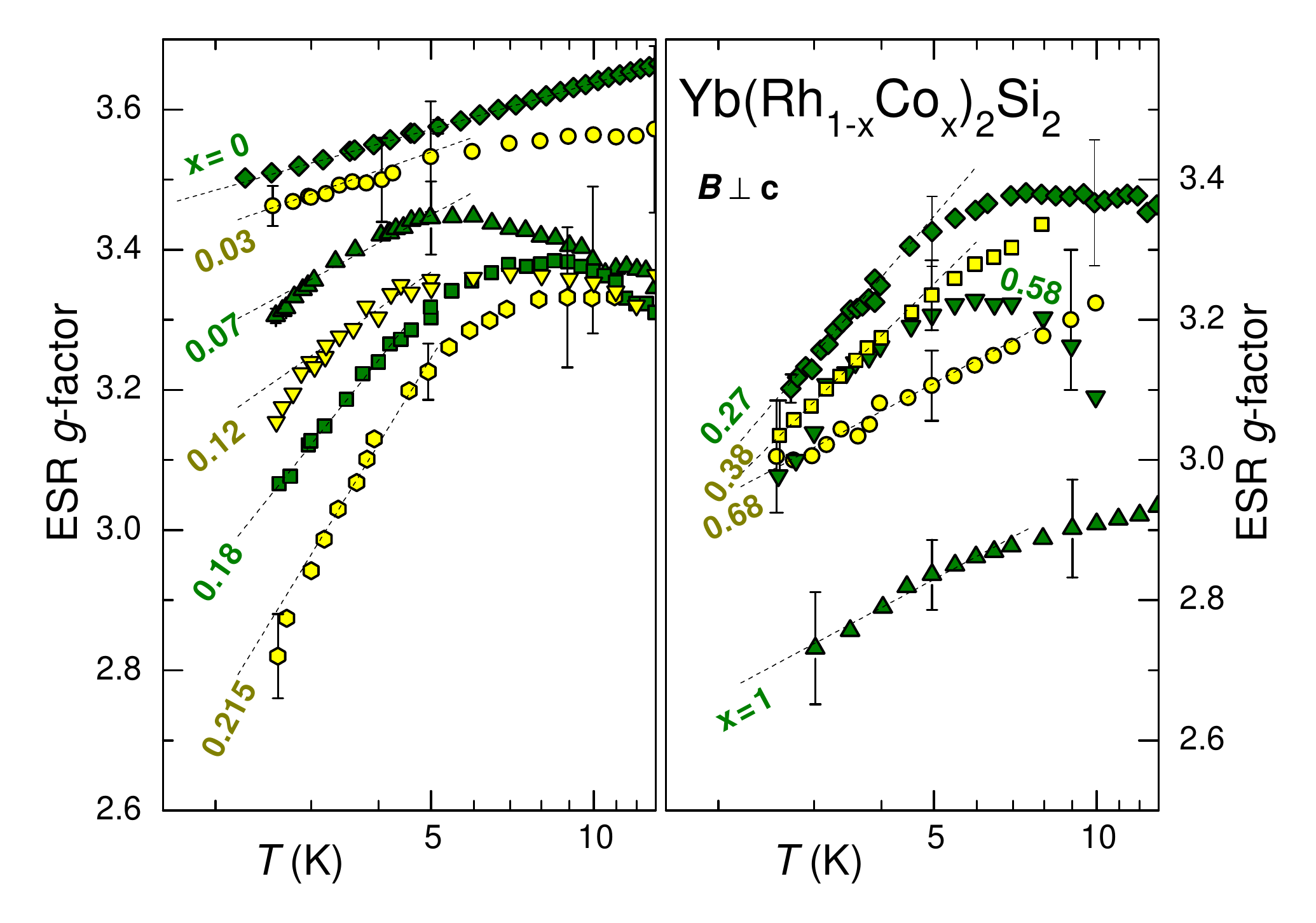}
  \caption{(color online) Temperature dependence of the ESR $g$-factor for the field orientation $\boldsymbol{B}\perp c$-axis. Dashed lines guide the eyes and emphasize the trend in the low temperature behavior.} 
 \label{gTemp}
\end{figure}

\indent The evolution of the linewidth and its temperature dependence is shown in Fig.~\ref{fig5}. Different characteristic behaviors suggest three temperature regimes: a linear temperature regime which separates a regime below $T\approx4$~K, 
%
%
characterized by tendencies to saturation or to low temperature broadening, from a regime above $T\approx 12$~K where an additional linewidth broadening indicates the influence of relaxation via the first excited CEF-level.\cite{sichelschmidt03a}
The linear linewidth behavior of the middle temperature regime reminds to a Korringa law of local-moment relaxation toward conduction electrons. This is displayed in Fig.\ref{fig5} by the linear dashed line for the $x=0.27$ sample which among all samples displays the largest temperature slope $b_{\perp}=\delta\Delta B_{\perp}/\delta T$ of 22~mT/K. The linearly extrapolated linewidth at zero-temperature has roughly the same value for the samples with $x\le0.27$. With further increasing $x$ $b_{\perp}$ decreases again and for $x\ge0.68$ is approximately the same as for $x=0.03$. 
In the temperature regime below $T\approx4$~K a deviation from linearity is observed for $0.07\le x\le0.27\;$ \cite{sichelschmidt10b} and $x\ge0.68$.
This is most obviously seen for the $x=1$ linewidth,  showing a considerable broadening above the magnetic ordering temperature. For $0.07\le x\le0.27$ and $x=0.68$ small positive curvatures in the low temperature linewidth behaviors rather seem to indicate a tendency for saturation. This tendency was discussed for $0.07\le x\le 0.27$ in terms of a relation between the zero-temperature linewidth to the residual electrical resistivity.\cite{sichelschmidt10b}\\ 
The behavior of the ESR $g$-factor is shown on a logarithmic temperature scale in Fig.~\ref{gTemp}. In general, as indicated by the dashed lines, all the $g$-data are characterized by a logarithmic downturn towards low temperatures. 
%
%
This is most nicely developed for $x=0$ where it was related to a small characteristic energy for the ground state Kramers doublet.\cite{kochelaev09a} Whereas in this low temperature region the temperature variation is largest for $x=0.215$ the absolute $g$-values steadily decrease with Co doping (see also Fig.~\ref{fig4}), indicating the effect of changes in the crystal electric field. The influence of the first excited crystal-field levels may be seen in the decrease of the $g$-factors at higher temperatures, being quite sensitive to the Co-content. Unfortunately, due to the insufficient experimental accuracy, further information on these excited levels could not be obtained.\\
The ESR intensity per Yb-ion does not change within experimental accuracy across the doping series. Previous detailed investigations of the number of ESR active Yb spins in YbRh$_{2}$Si$_{2}$ have shown that \textit{all} Yb ions contribute to the ESR signal.\cite{wykhoff07b} Hence, for Yb(Rh$_{\rm 1-x}$Co$\rm _x$)$_2$Si$_2$, where any evidence for Co magnetism is absent, the ESR properties are always determined by \textit{all} Yb ions despite the large variation of the Kondo interaction with Co-doping.\cite{klingner11a} 
%
%
%
%
\section{Discussion}
\label{disc}
\begin{figure}[t]
 \centering
 \includegraphics[width=0.8\columnwidth]{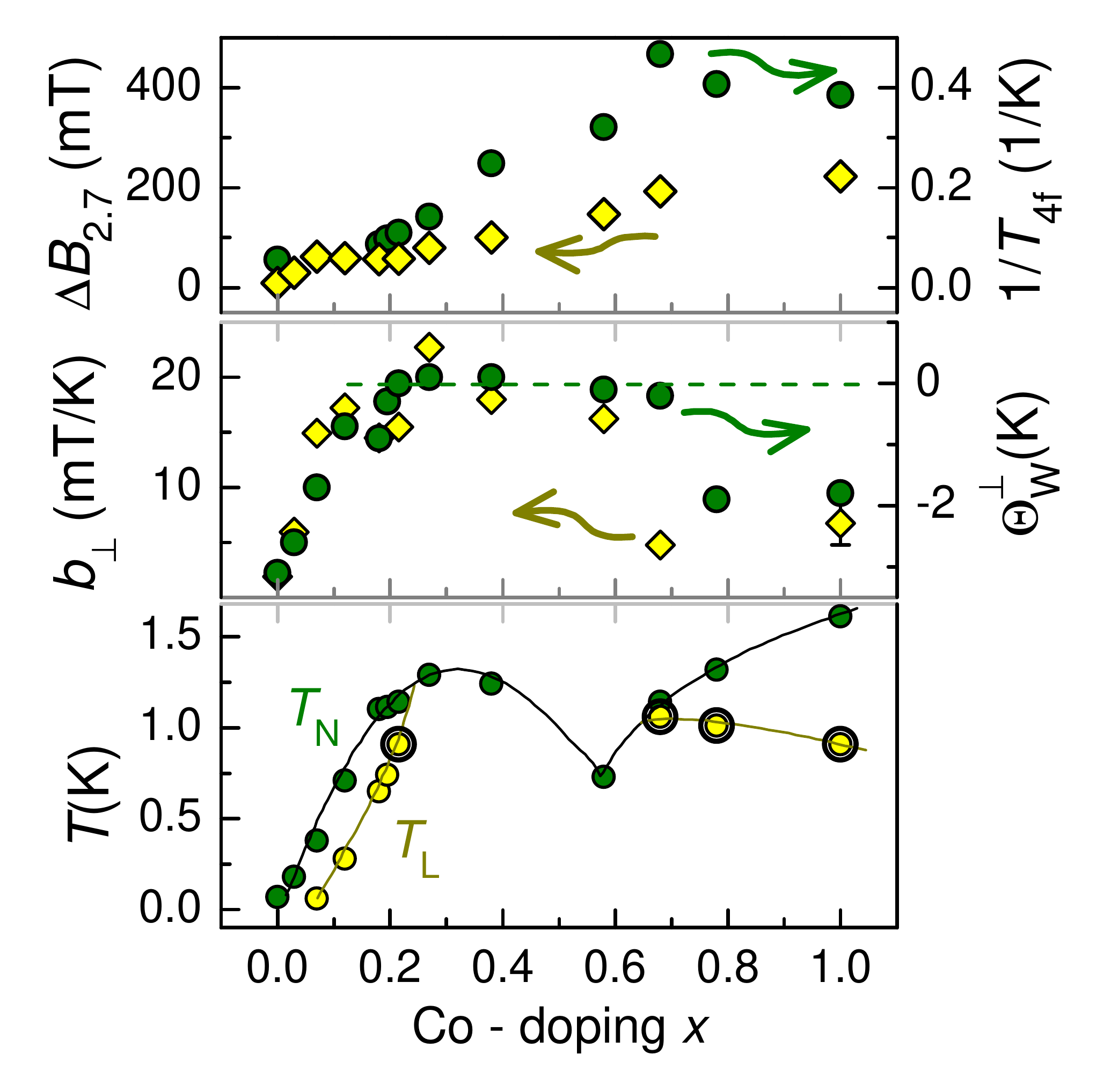}
  \caption{(color online) Comparison of ESR linewidth $\Delta B_{2.7}$ at $T=2.7$~K and linewidth temperature slope $b_{\perp}$ between 4 and 8~K (9 and 13~K; x=1) with the characteristic temperature $T_{4f}$ deduced from the $4f$- specific heat and the low-temperature, in-plane Weiss temperature $\Theta_{\rm W}^{\perp}$  of Yb(Rh$_{\rm 1-x}$Co$\rm _x$)$_2$Si$_2$ (data from Ref.\onlinecite{klingner11a}).
Bottom frame shows magnetic phase diagram \cite{klingner11a} as determined from anomalies of $\chi(T)$ and $C(T)$. Below temperatures $T_{\rm N}$ and $T_{\rm L}$  magnetic order appears to be incommensurate and commensurate, respectively. While the transition at $T_{\rm N}$ is always second-order, the transition at $T_{\rm L}$ is second order for $x\le0.22$ and first-order for $x>0.22$.} 
 \label{PD}
\end{figure}

Collecting information on the competition of the local Kondo and the intersite RKKY exchange interactions is essential for understanding the magnetic behavior of Yb(Rh$_{1-x}$Co$_x$)$_2$Si$_2$. The doping with Co enhances the size of the ordered moment and stabilizes the AFM order by weakening the Kondo interaction.\cite{klingner11a} The observed ESR parameters show dependencies on the Co-content which are sensitively related to the Kondo effect, to the anisotropy induced by the crystal field as well as to the presence and character of magnetic phase transitions below the investigated temperature range.
In the following this will be discussed by analyzing the linewidth parameters shown in Fig.\ref{PD} in terms of the relevant magnetic properties of Yb(Rh$_{\rm 1-x}$Co$\rm _x$)$_2$Si$_2$.
%

We start with discussing the relation between the linewidth and the characteristic temperature $T_{4f}$ deduced from the magnetic entropy which reflects the energy scale of all exchange interactions acting on the lowest CEF doublet.\cite{klingner11a} We have chosen the linewidth $\Delta B_{2.7}$ at $T=2.7$~K as one possible measure of the spin fluctuations which dominate the linewidth at low temperatures. As shown in the upper frame of Fig. \ref{PD} $\Delta B_{2.7}(x)$ and $1/T_{4f}(x)$ show a similar behavior. For low Co-contents this similarity suggests that a decrease of the Kondo energy scale $T_{\rm K}$ leads to an increase of the ESR linewidth because $T_{\rm K}$ dominates $T_{4f}$ for Co-concentrations $x\le0.38$. In this respect it is important to mention also the effect of anisotropy of the exchange interactions to the linewidth. It has been shown that a collective spin mode of Kondo ions and conduction electrons enables a narrow linewidth in a Kondo lattice despite a strong magnetic anisotropy.\cite{kochelaev09a} Hence, despite the anisotropy (as seen in the ratio of $g_{\perp}(x)/g_{\|}(x)$, Fig.~\ref{fig4}) is largest at $x=0$, there the linewidth reaches its smallest value, supporting the collective spin mode model.
Note, that with increasing $x$ the increasing contribution of disorder to the linewidth should be considered as well. A discrimination of Kondo- and disorder contributions was qualitatively accomplished by a comparison of the linewidth-resistivity relation of pressurized YbRh$_{2}$Si$_{2}$ and Yb(Rh$_{\rm 1-x}$Co$\rm _x$)$_2$Si$_2$ with $x\le0.18$.\cite{sichelschmidt10b} The result suggests that disorder is not the dominant contribution in $\Delta B_{2.7}(x)$.\\ 
At high Co-contents the ESR linewidth should be considered in terms of a strongly suppressed $T_{\rm K}$ because $T_{4f}$ gets dominated by RKKY exchange interactions for $x>0.6$. A strongly reduced Kondo interaction is established from photoemission spectroscopy: YbCo$_{2}$Si$_{2}$ (with $T_{\rm K}\ll 1$~K \cite{klingner11b}) shows a rather weak interaction between the 4$f$ and the valence states in comparison to YbRh$_{2}$Si$_{2}$.\cite{klingner11a} We discuss the role of the RKKY interactions for the linewidth in terms of ferromagnetic fluctuations which were shown to play an important role for the observability of a spin resonance in Kondo lattice systems.\cite{krellner08a}
The evolution of ferromagnetic fluctuations in Yb(Rh$_{\rm 1-x}$Co$\rm _x$)$_2$Si$_2$ are reflected, for instance, in the Weiss temperature $\Theta_{\rm W}^{\perp}$ characterizing the in-plane magnetic susceptibility at low temperatures. $\Theta_{\rm W}^{\perp}(x)$ shows a clear maximum around $x=0.3$, where it even changes sign and becomes positive, indicating the dominance of ferromagnetic exchange interactions. 
As documented in the middle frame of Fig.~\ref{PD} this behavior is well correlated with the in-plane linear temperature slope of the linewidth, $b_{\perp}(x)$. For low $x$ we interpret this correlation, similar to $\Delta B_{2.7}(x)$, by a decreasing Kondo interaction up to $x\approx0.2$.\\ 
At this point it is worth to note that in compounds with no Kondo lattice behavior (very small Kondo energy scales) such as YbRh \cite{krellner08a} or YbCo$_{2}$Si$_{2}$ narrow ESR lines emerge from ferromagnetic correlated 4$f$ spins whereas antiferromagnetic correlations as in Yb$_4$Rh$_7$Ge$_6$ \cite{krellner08a} lead to an ESR linewidth too broad to be observable. That means, for Yb(Rh$_{\rm 1-x}$Co$\rm _x$)$_2$Si$_2$ above $x\approx0.2$, the ferromagnetic correlations among the 4$f$ moments presumably start to control the linewidth, preventing $b_{\perp}(x)$ to get further increased which may be expected from an increasing influence of the RKKY interaction. Moreover, at high Co-contents, narrow lines are also allowed by an anisotropy $g_{\perp}(x)/g_{\|}(x)$ being much smaller than for low Co-contents. This is consistent with a constant or slightly decreasing (for $x>0.6$) relaxation rate, see Fig. \ref{fig4}. 
Again, these relations between the ESR linewidth and the interplay of Kondo effect, RKKY interactions and  anisotropy support the theoretical basis describing the spin dynamics in terms of a collective spin mode of Kondo ions and conduction electrons.\cite{kochelaev09a}\\  
Next, we consider the linewidth behavior towards low temperatures. There, broadening contributions from magnetic fluctuations become important even far above the magnetic transition temperature because the resonance relaxation is very sensitive to the effect of critical spin fluctuations.\cite{seehra75a} The type of magnetic order underlying the magnetic phase diagram in Fig. \ref{PD} depends on the Co-content. As a consequence, characteristic differences in the low-temperature ESR broadening can be identified for low and high Co concentrations, roughly separated by the pronounced minimum of $T_{\rm N}$ near $x=0.5$. As can be seen in Fig.~\ref{fig5} for almost all investigated Co concentrations the ESR line gets continuously narrowed down to the lowest accessible temperatures. A considerable ESR broadening toward low temperatures is only seen for $x=1$. However, the $x=0.68$ sample may also show a similar broadening at lower temperatures because the linewidth temperature slope $b_{\perp}$ is similar to the one for $x=1$ at $T>9$~K. The linewidth of the samples with Co concentrations $0.07\le x\le0.27$ show deviations from the temperature linearity below $T\approx4$~K.\\
\indent The sensitivity of the ESR linewidth to the type of magnetic order in Yb(Rh$_{\rm 1-x}$Co$\rm _x$)$_2$Si$_2$ strongly resembles the observations from a local Gd$^{3+}$ ESR in the heavy-fermion compound Ce(Cu$_{1-x}$Ni$_{x}$)$_{2}$Ge$_{2}$. There, the nature of magnetic order changes significantly from $x\le0.5$ (local moment type magnetism) to $x\ge0.5$ (itinerant heavy-fermion band magnetism).\cite{krug-von-nidda98a} The temperature dependence of the Gd$^{3+}$ linewidth could characterize the different ground states and, in particular, provide information about Ce$^{3+}$ spin-fluctuations and the scattering of conduction electrons at the Gd$^{3+}$ spins (for which, moreover, non-Fermi liquid behavior could be identified). The effect of Ce$^{3+}$ spin-fluctuations could be described by a $1/\sqrt{T}$ dependence. In the case of Yb(Rh$_{\rm 1-x}$Co$\rm _x$)$_2$Si$_2$ the Yb$^{3+}$ spin fluctuations lead to an increase of the linewidth, as clearly visible for the data of $x=1$. As shown by the solid line in Fig.\ref{fig5} a $1/\sqrt{T}$ dependence reasonably describes the data.\\
Previous ESR investigations of YbRh$_{2}$Si$_{2}$ under pressure and with Co-dopings $x\le0.18$ have shown a close relationship between the $g$-factor and the magnetic susceptibility \cite{sichelschmidt10b}, pointing out a typical property of a \textit{local} ESR probe. This feature was discussed in terms of a molecular field model for YbIr$_{2}$Si$_{2}$ in Ref. \onlinecite{gruner10a} and for YbCo$_{2}$Si$_{2}$ in Sec. \ref{sec3A}. That means that internal magnetic fields lead to an effective $g$-factor $g_{\rm eff}$ which is shifted from the ionic $g$-factor $g_0$ depending on a molecular-field parameter $\lambda$:
\begin{figure}[t]
 \centering
 \includegraphics[width=1\columnwidth]{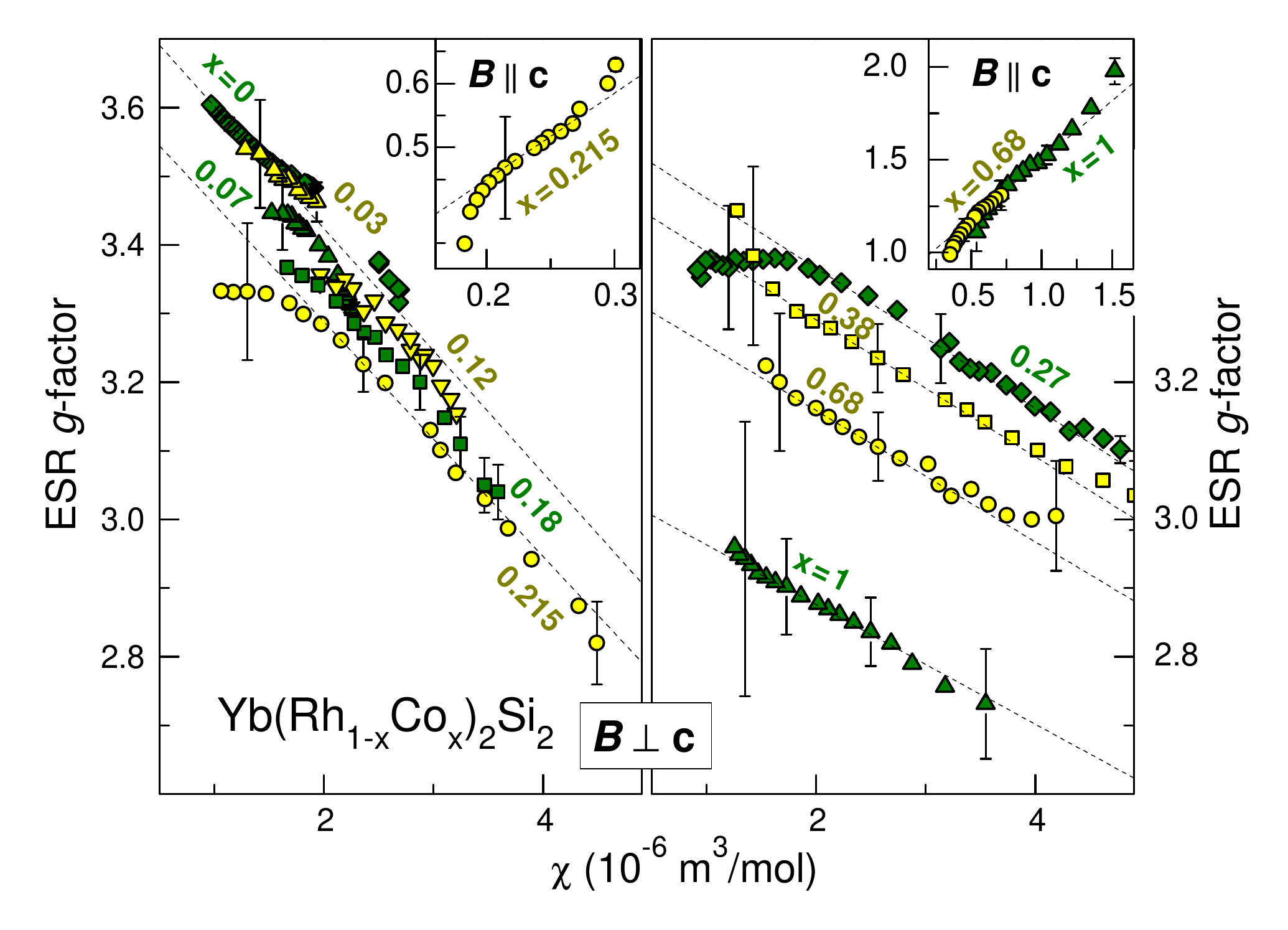}
  \caption{(color online) Relation between ESR $g$-factor and magnetic susceptibility $\chi$ with $\boldsymbol{B}\perp c$-axis. Insets show the relation for $\boldsymbol{B}\,\|\, c$-axis. Dashed lines indicate a molecular-field relationship as discussed in the text.} 
 \label{gChi}
\end{figure} 
\begin{table}[t]
\begin{center}
\begin{tabular}{c|c|c|c|c}
 $x$ & $g_0^{\perp}$ & $\lambda^{\perp}$ (kOe/$\mu_{\rm B}$) & $g_0^{\|}$ & $\lambda^{\|}$ (kOe/$\mu_{\rm B}$) \\ \hline
 0 & 3.78 & -3.3 & - & -\\
 0.215 & 3.63 & -3.3 & 0.18 & 525\\
 0.27 & 3.57 & -2 & - & -\\
 0.38 & 3.49 & -2 & - & -\\
 0.68 & 3.35 & -2 & 0.86 & 52\\
 1 & 3.05 & -2 & 0.86 & 52\\
\end{tabular}
\end{center}
\caption{Meanfield parameters of $g$ obtained from linear fits as shown by the dashed lines in Fig. \ref{gChi}.}
\label{table1}
\end{table}
%
%
\begin{equation}
g_{\rm eff}=g_0[1+\lambda\chi(T)] 
\end{equation}
For Yb(Rh$_{\rm 1-x}$Co$\rm _x$)$_2$Si$_2$ the validity of this equation is nicely established as shown by the dashed lines in Fig.~\ref{gChi} with the values listed in Table \ref{table1}. 
The values $g_{0}$ successively decrease with increasing Co-content. Discrepancies between these values and the values $g^{m}$ obtained from saturation magnetization could be due to Van-Vleck contributions to the susceptibility.\cite{kutuzov08a} The in-plane molecular-field parameter $\lambda^{\perp}$ which is sensitive to the interactions for the $x-y$ components of the Yb-moments shows negative values pointing to an antiferromagnetic exchange for the $x-y$ components.
An \textit{abrupt} change from $\lambda^{\perp} =-3.3\, {\rm kOe/\mu_{\rm B}}$ for $x\le 0.215$ to $\lambda^{\perp} =-2\, {\rm kOe/\mu_{\rm B}}$ for $x\ge 0.27$ could be related to a corresponding change in the magnetic structure. M\"ossbauer results on pure YbCo$_{2}$Si$_{2}$ indicate the ordered moment to be within the basal plane.\cite{hodges87a} The out-of plane internal fields lead to a ferromagnetic $g$-shift as shown in the insets of Fig.~\ref{gChi}. There, positive and large molecular-field values $\lambda^{\|}$ describe the data. The strongly enhanced value of $\lambda^{\|}$ for $x=0.215$ is a consequence of the small value of $g_0^{\|}$, since within a Heisenberg picture $\lambda$ can be related to the exchange parameter $J$ as $\lambda\propto J/g^2$. Hence, for $x\ge0.68$ $J^{\|}$ is twice as large than for $x=0.215$ supporting an increase of the ferromagnetic exchange with increasing $x$.  
These results suggest that the exchange for the z and the x-y components of the local moment are predominantly ferromagnetic and antiferromagnetic, respectively. 
\section{Conclusion}
The series Yb(Rh$_{\rm 1-x}$Co$\rm _x$)$_2$Si$_2$ show well defined ESR spectra with properties typical for local Yb$^{3+}$ spins. The ESR $g$-factor shows a distinct and uniform variation with the Co-content and no indications for a change in the symmetry of the ground state Kramers doublet could be found. 
The linewidth evolution with the Co-content clearly reflects variations of both the magnetic order and the Kondo interaction. 
A strong reduction of the latter by Co-doping leads to a line broadening. This demonstrates that the Kondo interaction crucially determines the ESR linewidth in Kondo lattices, especially in the presence of a strong magnetic anisotropy (i.e.~$g_{\perp}/g_{\|}$ is large) which is the case for Yb(Rh$_{\rm 1-x}$Co$\rm _x$)$_2$Si$_2$ with a small Co-content. Furthermore, our results suggest that ferromagnetic correlations may remain as the dominant mechanism for narrow ESR lines when the Kondo interaction is strongly suppressed and the magnetic anisotropy is sufficiently small, i.e., at high Co-contents of Yb(Rh$_{\rm 1-x}$Co$\rm _x$)$_2$Si$_2$. This is a new aspect of the previously stated relevance of ferromagnetic correlations among 4$f$ spins for a narrow ESR in Kondo lattice systems.\\ 
The character of the linewidth temperature dependence towards low temperatures ($T\lesssim4$~K) is clearly different for low and high Co-dopings. This indicates changes in the magnetic fluctuations upon Co-doping due to changes in the AFM structure. 
These changes could also be inferred from a change in the in-plane molecular-field around $x=0.22$ where the linearity between the magnetic susceptibility $\chi$ and the ESR $g$-values changes slope. Moreover, this slope shows a different sign for $g_{\|}(\chi)$ and $g_{\perp}(\chi)$ suggesting the exchange for the z and the x-y components of the moment to be predominantly ferromagnetic and antiferromagnetic, respectively. 
\section*{Acknowledgements}
We thank F. Garcia, Hans-Albrecht Krug von Nidda and B.I. Kochelaev for fruitful discussions and acknowledge the Volkswagen foundation (I/84689) for financial support.
%
%

%
%
\end{document}